# Predication of topological states in the allotropes of group-IV elements

Chengyong Zhong [1,2,3]

[1] *National Laboratory of Solid State Microstructures and Department of Physics, Nanjing University, Nanjing 210093, China*

[2] *Collaborative Innovation Center of Advanced Microstructures, Nanjing University, Nanjing 210093, China*

[3] *Institute for Advanced Study, Chengdu University, Chengdu 610106, China*

E-mail: zhongcy90@gmail.com

Three-dimensional (3D) topological insulators (TIs) have been studied for approximately fifteen years, but those made from group-IV elements, especially Ge and Sn, seem particularly attractive owing to their nontoxicity, sizable intrinsic spin–orbit coupling (SOC) strength and natural compatibility with the current semiconductor industry. However, group-IV elemental TIs have rarely been reported, except for the low temperature phase of $α$-Sn under strain. Here, based on first-principles calculations, we propose new allotropes of Ge and Sn, named T5-Ge/Sn, as desirable TIs. These new allotropes are also highly anisotropic Dirac semimetals if the SOC is turned off. To the best of our knowledge, T5-Ge/Sn are the first 3D allotropes of Ge/Sn that possess topological states in their equilibrium state at room temperature. Additionally, their isostructures of C and Si are metastable indirect and direct semiconductors. Our work not only reveals two promising TIs, but more profoundly, we justify the advantages of group-IV elements as topological quantum materials (TQMs) for fundamental research and potential practical applications, and thus reveal a new direction in the search for desirable TQMs.

# 1 Introduction

Topological quantum materials (TQMs), which include topological insulators (TIs) and topological semimetals (TMs), have received perennial interest in condensed matter physics and material science over the past decades [1-9]. TIs, as the first class of TQMs, have been recognized to potentially exhibit numerous applications in various fields, owing to their intriguing topological electronic properties, which include quantum computation and high-temperature superconductivity. Although TIs have only been studied for approximately fifteen years [3], many three-dimensional (3D) and two-dimensional (2D) TIs have already been experimentally confirmed and even more have been theoretically proposed [10-12]; however, desirable 3D TIs for application and especially those that are compatible with the current mainstream semiconductor industry are still rare. From the application point of view, a desirable TI would be nontoxic, easily manipulated and composed of widely used elements (e.g., Si, Ge, Sn). Unfortunately, the strong spin–orbit coupling (SOC) required by TIs generally limits the choice of materials to compounds containing heavy elements (e.g., the $Bi_2Se_3$ compounds [13], ternary Heusler compounds [14, 15]), complex heterostructures (e.g., a HgTe quantum well [16]), or poisonous elements (e.g., mercury). Therefore, it is necessary to discover more desirable TIs for the purpose of a more effective integration with conventional semiconductor devices.

Considering these reasons, the allotropes of group-IV elements (e.g., C, Si, Ge, Sn, and Pb) seem to be good alternatives in the search for desirable TIs for at least two reasons: first, they can be efficiently integrated with various current semiconductor devices and modulated by standard semiconductor techniques; second, their numerous metastable polymorphs are very beneficial to find potentially promising TIs. In fact, the study of TQMs containing group-IV elements started almost simultaneously when graphene was successfully discovered, as it is well known that graphene is the prototype of TIs or TMs depending on whether SOC is turned on or off [17, 18]. Because of the requirement of strong SOC in TIs, C and Si are not suitable for use in finding TIs, owing to their intrinsic weak SOC; for instance, the SOC-induced band gap in graphene and silicene is only 0.02 meV and 1.9 meV, respectively [19]. Toxic Pb is also not a good choice despite it having the strongest

SOC among the group-IV elements. Therefore, the most promising candidates are Ge and Sn.

However, the investigations on TIs composed of systems of Ge and Sn are far from complete. Only the graphene analogues of Ge and Sn (i.e., germanene and stanene) are 2D TIs and have successfully been grown on different substrates [20]. As for the 3D structures, 3D Ge with a diamond cubic structure is an indirect band gap semiconductor at ambient conditions [21]. Although many 3D Ge polymorphs have also been proposed using first-principles calculations or realized in experiments [22-27], they are usually trivial semiconductors and none of them is a TI. The diamond cubic structure of Sn, known as $\alpha$-Sn, is a low-temperature phase of Sn (stable below 13.2 °C). In contrast to its much more common metallic allotrope $\beta$-Sn, $\alpha$-Sn is a zero-bandgap semiconductor with band inversion. Indeed, $\alpha$-Sn has received much attention as it can be engineered into different topological states, that is, a TI, Dirac semimetal and Weyl semimetal in the presence of external strains, magnetic fields and circularly polarized light [28-32]. Nevertheless, to the best of our knowledge, strained $\alpha$-Sn is the only elemental 3D TI composed from group-IV elements. Given the advantages and incompleteness of TIs in group-IV elements, identifying new group-IV elemental 3D TIs that are stable at room temperature is of great importance for both fundamental research and potential practical applications.

In this paper, based on first-principles calculations, we report new metastable 3D allotropes of Ge and Sn with topological electronic properties, named T5-Ge and T5-Sn. Their stabilities are thoroughly investigated through lattice dynamics and *ab initio* molecular dynamics (AIMD). Without considering SOC, T5-Ge and T5-Sn are highly anisotropic 3D Dirac semimetals; correspondingly, they are TIs with a negative bandgap of −21 meV and indirect bandgap of 31 meV if the SOC is switched on. Their topological properties are confirmed by calculation of the topological indices and topological surface states. To the best of our knowledge, T5-Ge/Sn are the first 3D allotropes of Ge/Sn that possess topological states in their equilibrium states at room temperature. Moreover, considering that group-IV elements usually share the same atomic structure, their C and Si isostructures are also studied. It is observed that T5-C and T5-Si are an indirect bandgap semiconductor and a direct bandgap semiconductor, respectively. Our findings not only propose the two 3D structures with topological states, but most significantly, we propose that the TIs in group-IV elements are nontoxic, easily manufactured and more compatible with the current semiconductor industry and thus present a new direction in the search for desirable TIs.

## 2 Results and discussion

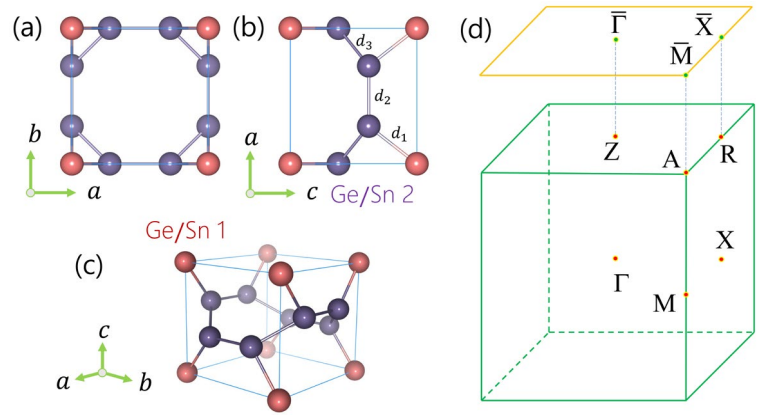

**Fig. 1 (a)** Top view, **(b)** side view and **(c)** perspective view of T5-Ge/Sn. Two irreducible atoms are colored in dark red (Ge/Sn 1) and deep purple (Ge/Sn 2). Three different bonds are indicated by $d_1$, $d_2$ and $d_3$. **(d)** The 3D BZ of the tetragonal lattice, and its 2D BZ projected onto the (001) surface. The high-symmetry $k$ points are labeled.

As shown in Fig. 1 (a–c), T5-Ge and T5-Sn share the same tetragonal crystal system with space group P-4M2 (No. 115) and contain five atoms in the unit cell. One of the atoms occupies the vertex of the cuboid, which corresponds to the *1a* (0,0,0) Wyckoff position, and the other four atoms distributed at two equivalent lateral facets are related by S$_4$ rotoinversion symmetry along the tetragonal axis (defined as the *c* axis), which represents the *4j* (0, 0.2776, 0.3803) Wyckoff site, and forms a buckled square-octagon lattice in the *a-b* plane [see Fig. 1(a)]. The corresponding first Brillouin Zone (BZ) with high symmetry points is shown in Fig. 1(d). After full relaxation of the structures, the optimized lattice constants are *a* = 5.63/6.49 Å and *c* = 5.23/5.99 Å for T5-Ge/Sn. According to symmetry, there are three kinds of bonds in T5-Ge/Sn; the first is between Ge/Sn 1 and Ge/Sn 2 ($d_1$ = 2.53/2.91 Å), the second is between two Ge/Sn 2 atoms within one facet ($d_2$ = 2.50/2.89 Å), and the last bond is that between two nearby Ge/Sn 2 atoms belonging to two neighboring facets ($d_3$= 2.54/2.92 Å), which are presented in Table 1 along with other crystalline information.

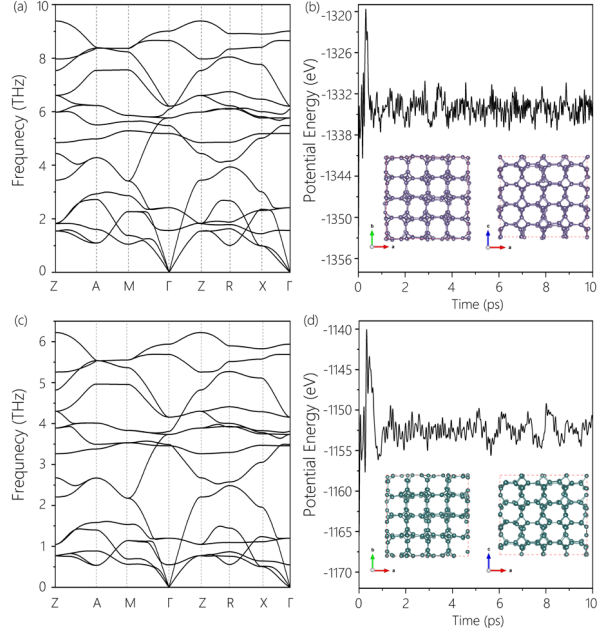

**Fig. 2 (a, c)** Phonon spectra of T5-Ge/Sn. **(b, d)** Total potential energy fluctuation of T5-Ge/Sn during the AIMD simulation at 300 K. The insets are snapshots after 10 ps simulations.

**Table 1** Lattice constants $a/c$ (Å), bond lengths $d_1/d_2/d_3$ (Å), elastic constants (GPa) and band gap (eV) from HSE06 (PBE).

|       | a    | c    | $d_1$ | $d_2$ | $d_3$ | C11 | C33 | C44 | C66 | C12 | C13 | $E_g$       |
|-------|------|------|-------|-------|-------|-----|-----|-----|-----|-----|-----|-------------|
| T5-C  | 3.42 | 3.23 | 1.54  | 1.50  | 1.59  | 793 | 820 | 139 | 84  | 16  | 55  | 2.83(1.82)  |
| T5-Si | 5.25 | 4.94 | 2.35  | 2.37  | 2.38  | 132 | 133 | 14  | 7   | 16  | 24  | 1.58(0.99)  |
| T5-Ge | 5.63 | 5.23 | 2.50  | 2.53  | 2.54  | 88  | 88  | 7   | 4   | 3   | 12  | 0           |
| T5-Sn | 6.49 | 5.99 | 2.89  | 2.91  | 2.93  | 47  | 44  | 3   | 2   | 5   | 12  | 0           |

Stability examination is crucial to prove the existence of the theoretically proposed phase. Here, the dynamic, thermodynamic and mechanical stabilities of T5-Ge/Sn are thoroughly assessed through state-of-the-art theoretical techniques. Their phonon spectra show there are no soft modes throughout the entire BZ, which demonstrates that they are dynamically stable [see Fig. 2(a) and (c)]. The thermodynamic stability of T5-Ge/Sn is verified through the AIMD simulations at 300 K [see Fig. 2(b) and (d)] in the framework of the NVT ensemble. One can observe that their structures are well maintained after 10 ps, with a time step of 1 fs. For the tetragonal crystal phase, the independent elastic constants are C11 = 88 GPa / 47 GPa, C33 = 88 GPa / 44 GPa, C44 = 7 GPa / 3 GPa, C66 = 4 GPa / 2 GPa, C12 = 3 GPa / 5 GPa, C13 = 12 GPa / 12 GPa for T5-Ge / T5-Sn, respectively (see Table 1). Obviously, these results meet the mechanical stability criteria for tetragonal systems [33]: C11 > 0,

C33 > 0, C44 >0, C66 >0, (C11 − C12) >0, (C11 + C33 − 2C13) >0, [2(C11 + C12) + C33+4C13] >0, which confirms that they are also mechanically stable. We also present their formation energies in comparison with those of diamond structures in Table S1, which shows that they are metastable group-IV elemental phases.

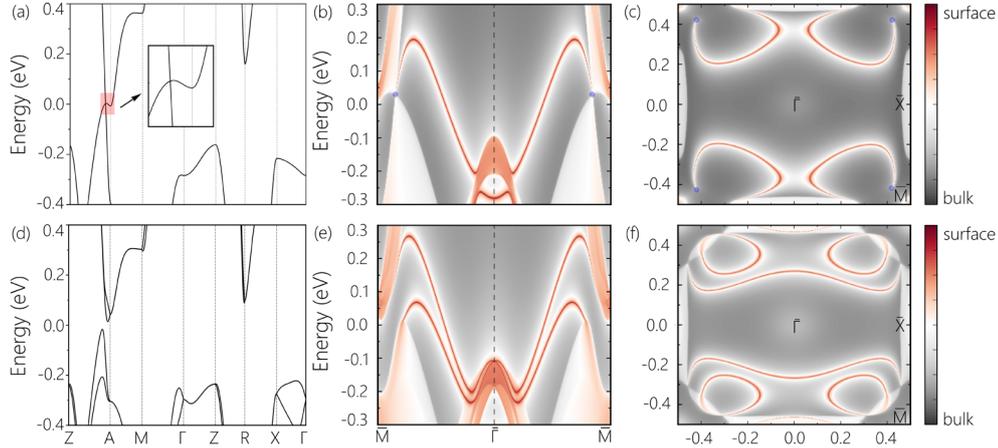

**Fig. 3 (a)** HSE band structure of T5-Sn without SOC. The inset is the amplified band structure corresponding to the part shadowed in red. **(b)** Band structure of T5-Sn projected onto the (001) surface. **(c)** Surface states in the 2D BZ of the (001) surface at a fixed energy crossing the projected 2D Dirac points. The blue dots in (b, c) indicate the locations of the projected 2D Dirac points. **(d-f)** Results of T5-Sn with SOC corresponding to (a–c), where the surface states in (f) are calculated at a fixed energy of the Fermi level.

Noticing the similarity of the topological electronic properties between T5-Ge and T5-Sn, as well as the more pronounced SOC effect of T5-Sn, here, we mainly focus on the topological electronic properties of T5-Sn, and those of T5-Ge will be briefly discussed later. Without a special statement, the electronic properties below are calculated with the HSE06 functional. The calculated electronic band structure of T5-Sn without including SOC around the Fermi level is shown in Fig. 3(a). One observes that the valence band and conduction band cross each other to form a point along Z-A. Interestingly, one Dirac band is nearly linear but the other is parabolic-like around the Dirac point, which implies that the Dirac cone is highly anisotropic [see the inset of Fig. 3(a)]. Because we do not include the degree of spin, each band is doubly degenerate, the crossed point should be quadruple degenerate, thus T5-Sn is a 3D Dirac semimetal. Given the nontrivial topology of a 3D Dirac semimetal, the projected 2D Dirac points and Fermi arcs are expected to be observed on some surfaces of T5-Sn.

As shown in Fig. 3(b) and (c), a pair of Dirac points (colored in blue) can be clearly observed at the electronic band structure of the (001) surface along $\bar{M} - \bar{\Gamma}$ [see Fig. 3(b)] and the Fermi arcs across the BZ boundary can also be clearly seen between them [see Fig. 3(c)]. Furthermore, there are no extraneous bands that pass through the Fermi level [see Fig. 3(a)], except the Dirac bands, and no extraneous trivial surface states are entangled with the nontrivial ones [see Fig. 3(b) and (c)], which is very beneficial for detecting the Dirac point and its topological nontrivial surface states in experiments.

It is known that the SOC strength of Sn is sufficient to induce observable physical properties at ambient conditions; for instance, the Dirac-point gap by SOC in stanene is approximately 101 meV [19]. Similarly, the Dirac-point gap of T5-Sn is 31 meV after including SOC [see Fig. 3(d)]. Using the topological quantum chemistry method proposed by Bradlyn *et al.* [34], we verified that T5-Sn is a TI with topological index $Z_2$=1, which belongs to the strong topological class 1. The appearance of the topologically guaranteed surface states is one of the most significant consequences of TIs. To further unveil the topological nature of T5-Sn, we calculate its electronic states on the (001) surface, from which, the Dirac cone formed at the $\bar{\Gamma}$ point can be clearly observed [see Fig. 3(e)]. Fixed at the Fermi level, its isoenergy surface of the projected surface states on the 2D BZ of the (001) surface distinctly shows that only the two topological surface bands exist around the Fermi level, which is very helpful for experimentally finding the topological surface states.

Strain engineering has been a ubiquitous paradigm to tailor the electronic band structure and harness the associated new or enhanced fundamental properties in TQMs, just as *α*-Sn demonstrates different topological states under compressive or tensile strain [30]. From the stress-strain relationship, we find that T5-Sn can sustain at least 2% uniaxial strain along the *a* and *c* directions [see Fig. 4(a)]. The two kinds of strains have a different impact on their topological electronic properties. Without including SOC, the Dirac points are ruined after applying uniaxial strain along the *a* direction while they remain intact along the *c* direction [see Fig. 4(b–e)]. The $S_4$ rotoinversion symmetry would be destroyed when the strain is imposed along the *a* direction, whereas the symmetry would remain unchanged along the *c* direction. Therefore, the Dirac semimetal of T5-Sn is protected by $S_4$ rotoinversion symmetry. In addition, we calculate the topological surface states of T5-Sn under 2% compressive and tensile strains along the *c* direction, which confirms that T5-Sn is still a TI under strain along the *c* direction [see Fig. 4(f) and (g), their phonon spectra are shown in Fig. S1].

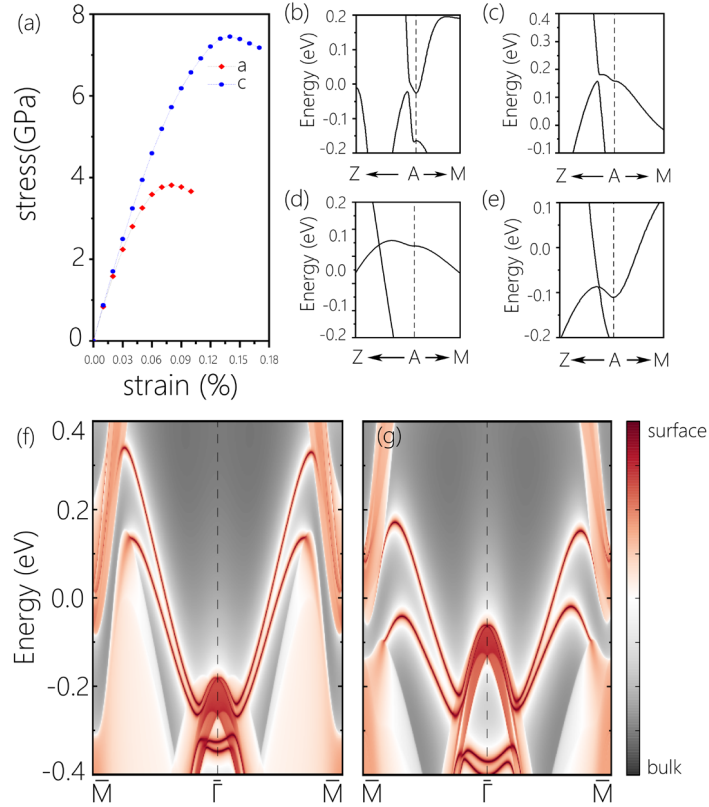

**Fig. 4 (a)** Stress-strain relationship for T5-Sn along the uniaxial *a* and *c* directions. The band structures of T5-Sn under **(b, d)** 2% compressive strain and **(c, e)** 2% tensile strain along the uniaxial *a/c* directions. The topological surface states of T5-Sn under **(f)** 2% compressive strain and **(g)** 2% tensile strain.

Despite the topological electronic properties of T5-Ge being similar to those of T5-Sn, they still demonstrate a few small differences. For example, without including SOC, T5-Ge is an even more notable anisotropic Dirac semimetal, of which one band is nearly flat but the other is linear [see Fig. 5(a)]; and more interestingly, a negative band gap of 21 meV appears after switching SOC on [see Fig. 5(b)]. As a 3D Dirac semimetal, we also calculate its surface states and Fermi arc to confirm the topology [see Fig. 5(c)], and the topological surface states of T5-Ge as a TI are given as well [see Fig. 5(d)].

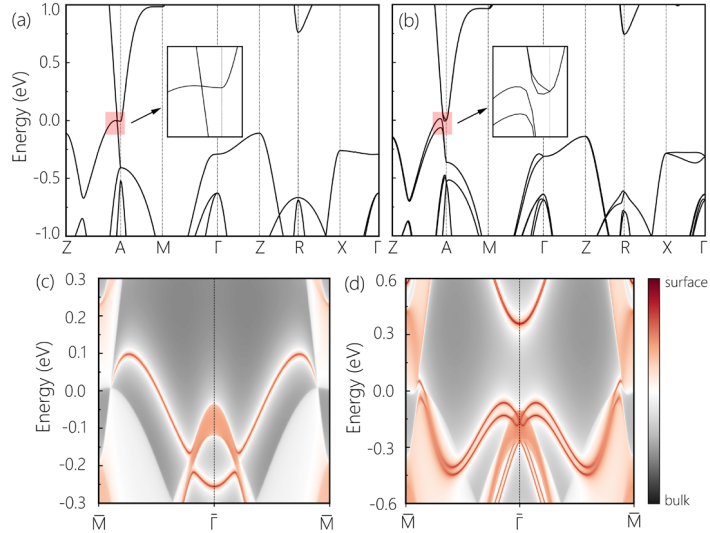

**Fig. 5** HSE band structure of T5-Ge **(a)** without SOC and **(b)** with SOC. The inset is the amplified band structures corresponding to the parts shadowed in red. Band structure of T5-Sn projected onto the (001) surface **(c)** without SOC and **(d)** with SOC.

Before concluding, we would like to provide a few remarks about our work. Group-IV elements have an $s^2p^2$ valence electronic configuration, which usually leads to common crystalline structures and similar chemical features, but can also lead to significant differences. Here, we also studied the T5 isostructures of C/Si and found that they are metastable phases, which is confirmed by the calculation of their phonon spectra and independent elastic constants [see Fig. 6 (a, b) and Table 1]. In contrast, T5-C is a normal semiconductor with an HSE (PBE) indirect bandgap of 2.83 eV (1.82 eV), while T5-Si is a semiconductor with a direct bandgap of 1.58 eV (0.99 eV), as shown in Fig. 6 (c, d). The calculated optical absorption of T5-Si as well as the air mass 1.5 solar spectral irradiance (see Fig. S2) indicate the potential of T5-Si as an absorber of sunlight in a photovoltaic device.

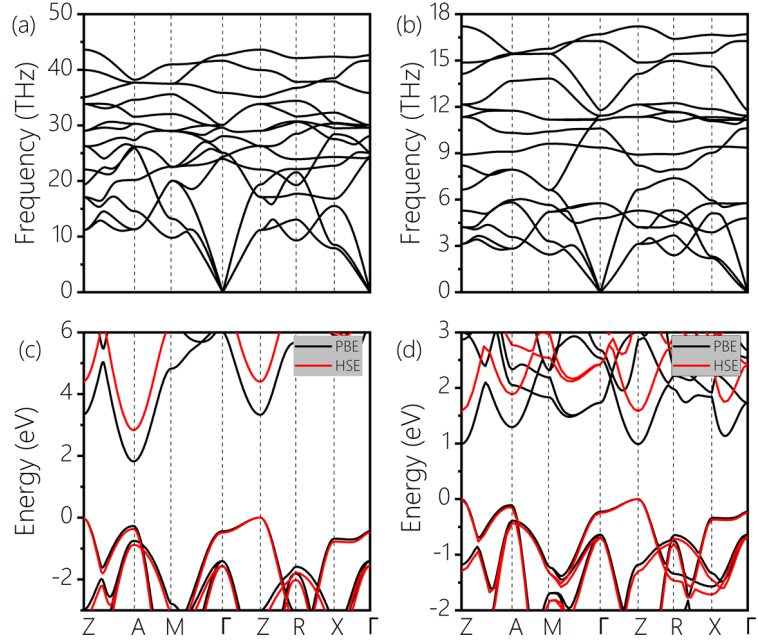

**Fig. 6** Phonon spectra of **(a)** T5-C and **(b)** T5-Si. The band structures of **(c)** T5-C and **(d)** T5-Si under the PBE and HSE levels.

In addition, we would like to point out that the possibility of alloying different group-IV elements to optimize the material properties has been experimentally considered for many years, such as SiGe or GeSn alloys for better optoelectronic performance [35, 36]. Therefore, the implication of alloys of group-IV elements with T5 structures on the electronic, optical and topological properties is very worthy of being studied in the future.

Finally, we briefly discuss the TQMs in group-IV elements (e.g., C, Si, Ge and Sn). The most distinct features for group-IV elements as TQMs are they are nontoxic, easily manipulated and compatible with current semiconductor devices. The increasing strength of SOC from C to Sn propels them to fit different kinds of TQMs. Given the strong SOC requirement of TIs, Ge and Sn are more suitable than C and Si to induce topological insulating states, as discussed in this work; inversely, the negligible SOC in C and Si makes them better candidates as TMs because the SOC-induced band inversion mechanism in TIs is not mandatory for TMs [37, 38]. In fact, many TMs, including nodal-line semimetals or nodal-surface semimetals, have been reported in carbon or silicon allotropes [39-45]. More recently, even 2D higher-order TIs have been claimed in 2D carbon materials (i.e., in graphdiyne) [46, 47], which is beyond the paradigm of SOC-related topological states discussed here. Thus, we are deeply convinced that group-IV elements can spawn numerous desirable and novel TQMs.

## 3 Conclusions

In summary, through first-principles calculations, we reveal a new 3D metastable crystalline form with a tetragonal space group (named as T5), which can be adopted by all group-IV elements. Among them, T5-C and T5-Si are indirect and direct bandgap semiconductors, while T5-Ge and T5-Sn are 3D Dirac semimetals or elemental TIs depending on whether the SOC is switch off or on. T5-Ge is the first 3D Ge allotrope possessing topological states and T5-Sn is the first 3D Sn allotrope that has topological states at ambient temperature and can sustain at least a 2% external uniaxial strain along the *c* axis. Meanwhile, the topological electronic properties are intrinsic to T5-Sn, compared with those of *α*-Sn that should be induced by external methods. In conclusion, we expect that our work of discovering TIs in nontoxic, easily manipulated and widely used group-IV elements can bring a fresh perspective to the search for TQMs.

## 4 Computational methods

We performed our first-principles calculations within the density functional theory formalism as implemented in the Quantum ESPRESSO (QE) code [48, 49]. The electron-ion interaction and the exchange-correlation functional between the valence electrons are described, respectively, by the projector augmented wave (PAW) and the generalized gradient approximation (GGA) parametrized by the Perdew-Burke-Ernzerhof (PBE) approach [50]. The kinetic energy cutoffs of 200 eV, 200 eV, 300 eV and 500 eV were employed for C, Si, Ge and Sn, respectively. The atomic positions were optimized until the residual forces converged at $10^{-2}$ eV/Å. The BZ was sampled with 8×8×8 k-point meshes in the scheme of Monkhorst–Pack [51]. Hybrid functional calculations of HSE06 [52] were mainly used to obtain the electronic properties. Phonon spectra were calculated using Phonopy [53] interfaced with QE. The topological surface properties were computed with WannierTools [54], of which the tight-binding Hamiltonian was generated by the maximally localized Wannier functions [55].

# Acknowledgements

The author is grateful for the helpful discussion with Pan Zhou. This project was funded by China Postdoctoral Science Foundation (Grant No. 2019M660107) the National Natural Science Foundation of China (Grant No. 11804039).

# References

1. M. Z. Hasan; C. L. Kane, Colloquium: Topological insulators. *Rev. Mod. Phys.* 82(4), 3045-3067 (2010)

2. N. P. Armitage; E. J. Mele; A. Vishwanath, Weyl and Dirac semimetals in three-dimensional solids. *Rev. Mod. Phys.* 90(1), 015001 (2018)

3. P. Liu; J. R. Williams; J. J. Cha, Topological nanomaterials. *Nat. Rev. Mater.* 4(7), 479-496 (2019)

4. C. N. Lau; F. Xia; L. Cao, Emergent quantum materials. *Mrs. Bull.* 45(5), 340-347 (2020)

5. Q. Niu, Advances on topological materials. *Front. Phys.* 15(4), 43601 (2020)

6. Y. X. Zhao, Equivariant PT-symmetric real Chern insulators. *Front. Phys.* 15(1), 13603 (2019)

7. Y. X. Zhao; Z. D. Wang, Novel Z(2) Topological Metals and Semimetals. *Phys. Rev. Lett.* 116(1), 016401 (2016)

8. Y. X. Zhao; A. P. Schnyder, Nonsymmorphic symmetry-required band crossings in topological semimetals. *Phys. Rev. B* 94(19), (2016)

9. F. Giustino, et al., The 2021 quantum materials roadmap. *J. Phys. Mater.* 3(4), (2021)

10. F. Tang; H. C. Po; A. Vishwanath; X. Wan, Comprehensive search for topological materials using symmetry indicators. *Nature* 566(7745), 486-489 (2019)

11. M. G. Vergniory; L. Elcoro; C. Felser; N. Regnault; B. A. Bernevig; Z. Wang, A complete catalogue of high-quality topological materials. *Nature* 566(7745), 480-485 (2019)

12. T. Zhang; Y. Jiang; Z. Song; H. Huang; Y. He; Z. Fang; H. Weng; C. Fang, Catalogue of topological electronic materials. *Nature* 566(7745), 475-479 (2019)

13. H. Zhang; C.-X. Liu; X.-L. Qi; X. Dai; Z. Fang; S.-C. Zhang, Topological insulators in Bi2Se3, Bi2Te3 and Sb2Te3 with a single Dirac cone on the surface. *Nat. Phys.* 5(6), 438-442 (2009)

14. H. Lin; L. A. Wray; Y. Xia; S. Xu; S. Jia; R. J. Cava; A. Bansil; M. Z. Hasan, Half-Heusler ternary compounds as new multifunctional experimental platforms for topological quantum phenomena. *Nat. Mater.* 9(7), 546-9 (2010)

15.  D. Xiao; Y. Yao; W. Feng; J. Wen; W. Zhu; X. Q. Chen; G. M. Stocks; Z. Zhang, Half-Heusler compounds as a new class of three-dimensional topological insulators. *Phys. Rev. Lett.* 105(9), 096404 (2010)

16.  B. A. Bernevig; T. L. Hughes; S. C. Zhang, Quantum spin Hall effect and topological phase transition in HgTe quantum wells. *Science* 314(5806), 1757-61 (2006)

17.  C. L. Kane; E. J. Mele, Quantum spin Hall effect in graphene. *Phys. Rev. Lett.* 95(22), 226801 (2005)

18. K. S. Novoselov; A. K. Geim; S. V. Morozov; D. Jiang; M. I. Katsnelson; I. V. Grigorieva; S. V. Dubonos; A. A. Firsov, Two-dimensional gas of massless Dirac fermions in graphene. *Nature* 438(7065), 197-200 (2005)

19.  S. Balendhran; S. Walia; H. Nili; S. Sriram; M. Bhaskaran, Elemental analogues of graphene: silicene, germanene, stanene, and phosphorene. *Small* 11(6), 640-52 (2015)

20.  A. J. Mannix; B. Kiraly; M. C. Hersam; N. P. Guisinger, Synthesis and chemistry of elemental 2D materials. *Nat. Rev. Chem* 1(0014 (2017)

21.  J. R. Chelikowsky; M. L. Cohen, Nonlocal pseudopotential calculations for the electronic structure of eleven diamond and zinc-blende semiconductors. *Phys. Rev. B* 14(2), 556-582 (1976)

22.  F. Kiefer; V. Hlukhyy; A. J. Karttunen; T. F. Fässler; C. Gold; E.-W. Scheidt; W. Scherer; J. Nylén; U. Häussermann, Synthesis, structure, and electronic properties of 4H-germanium. *J. Mater. Chem.* 20(9),  (2010)

23.  B. C. Johnson; B. Haberl; S. Deshmukh; B. D. Malone; M. L. Cohen; J. C. McCallum; J. S. Williams; J. E. Bradby, Evidence for the R8 phase of germanium. *Phys. Rev. Lett.* 110(8), 085502 (2013)

24. A. Mujica; C. J. Pickard; R. J. Needs, Low-energy tetrahedral polymorphs of carbon, silicon, and germanium. *Phys. Rev. B* 91(21), 214104 (2015)

25.  Z. Zhao; H. Zhang; D. Y. Kim; W. Hu; E. S. Bullock; T. A. Strobel, Properties of the exotic metastable ST12 germanium allotrope. *Nat. Commun.* 8(13909 (2017)

26.  Z. Tang; A. P. Litvinchuk; M. Gooch; A. M. Guloy, Narrow Gap Semiconducting Germanium Allotrope from the Oxidation of a Layered Zintl Phase in Ionic Liquids. *J. Am. Chem. Soc.* 140(22), 6785-6788 (2018)

27.  C. He; X. Shi; S. J. Clark; J. Li; C. J. Pickard; T. Ouyang; C. Zhang; C. Tang; J. Zhong, Complex Low Energy Tetrahedral Polymorphs of Group IV Elements from First Principles. *Phys. Rev. Lett.* 121(17), 175701 (2018)

28.  A. Barfuss, et al., Elemental topological insulator with tunable Fermi level: strained alpha-Sn on InSb(001). *Phys. Rev. Lett.* 111(15), 157205 (2013)

29. C. Z. Xu, et al., Elemental Topological Dirac Semimetal: alpha-Sn on InSb(111). *Phys. Rev. Lett.* 118(14), 146402 (2017)

30. D. Zhang; H. Wang; J. Ruan; G. Yao; H. Zhang, Engineering topological phases in the Luttinger semimetal α-Sn. *Phys. Rev. B* 97(19), 195139 (2018)

31. Q. Barbedienne, et al., Angular-resolved photoemission electron spectroscopy and transport studies of the elemental topological insulator α-Sn. *Phys. Rev. B* 98(19), 195445 (2018)

32. I. Madarevic, et al., Structural and electronic properties of the pure and stable elemental 3D topological Dirac semimetal α-Sn. *APL Mater.* 8(3), 031114 (2020)

33. J. F. Nye, *Physical properties of crystals*. Clarendon Press: 1985.

34. B. Bradlyn; L. Elcoro; J. Cano; M. G. Vergniory; Z. Wang; C. Felser; M. I. Aroyo; B. A. Bernevig, Topological quantum chemistry. *Nature* 547(7663), 298-305 (2017)

35. J. Doherty; S. Biswas; E. Galluccio; C. A. Broderick; A. Garcia-Gil; R. Duffy; E. P. O'Reilly; J. D. Holmes, Progress on Germanium–Tin Nanoscale Alloys. *Chem. Mater.* 32(11), 4383-4408 (2020)

36. J. Wagner; M. Núñez-Valdez, Ab initio study of band gap properties in metastable BC8/ST12 $Si_xGe_{1-x}$ alloys. *Appl. Phys. Lett.* 117(3), 032105 (2020)

37. A. P. Schnyder; S. Ryu; A. Furusaki; A. W. W. Ludwig, Classification of topological insulators and superconductors in three spatial dimensions. *Phys. Rev. B* 78(19), 195125 (2008)

38. C. Zhong; W. Wu; J. He; G. Ding; Y. Liu; D. Li; S. A. Yang; G. Zhang, Two-dimensional honeycomb borophene oxide: strong anisotropy and nodal loop transformation. *Nanoscale* 11(5), 2468-2475 (2019)

39. Y. Chen; Y. Xie; S. A. Yang; H. Pan; F. Zhang; M. L. Cohen; S. Zhang, Nanostructured Carbon Allotropes with Weyl-like Loops and Points. *Nano Lett.* 15(10), 6974-8 (2015)

40. H. Weng; Y. Liang; Q. Xu; R. Yu; Z. Fang; X. Dai; Y. Kawazoe, Topological node-line semimetal in three-dimensional graphene networks. *Phys. Rev. B* 92(4), 045108 (2015)

41. C. Zhong; Y. Chen; Y. Xie; S. A. Yang; M. L. Cohen; S. B. Zhang, Towards three-dimensional Weyl-surface semimetals in graphene networks. *Nanoscale* 8(13), 7232-9 (2016)

42. C. Zhong; Y. Chen; Z. M. Yu; Y. Xie; H. Wang; S. A. Yang; S. Zhang, Three-dimensional Pentagon Carbon with a genesis of emergent fermions. *Nat. Commun.* 8(15641 (2017)

43. W. Wu; Y. Liu; S. Li; C. Zhong; Z.-M. Yu; X.-L. Sheng; Y. X. Zhao; S. A. Yang, Nodal surface semimetals: Theory and material realization. *Phys. Rev. B* 97(11), 115125 (2018)

44. Z. Liu; H. Xin; L. Fu; Y. Liu; T. Song; X. Cui; G. Zhao; J. Zhao, All-Silicon Topological Semimetals with Closed Nodal Line. *J. Phys. Chem. Lett.* 10(2), 244-250 (2019)

45. S. Z. Chen; S. Li; Y. Chen; W. Duan, Nodal Flexible-surface Semimetals: Case of Carbon

Nanotube Networks. *Nano Lett.* 20(7), 5400-5407 (2020)

46. B. Liu; G. Zhao; Z. Liu; Z. F. Wang, Two-Dimensional Quadrupole Topological Insulator in gamma-Graphyne. *Nano Lett.* 19(9), 6492-6497 (2019)

47. X. L. Sheng; C. Chen; H. Liu; Z. Chen; Z. M. Yu; Y. X. Zhao; S. A. Yang, Two-Dimensional Second-Order Topological Insulator in Graphdiyne. *Phys. Rev. Lett.* 123(25), 256402 (2019)

48. P. Giannozzi, et al., QUANTUM ESPRESSO: a modular and open-source software project for quantum simulations of materials. *J. Phys.: Condens. Matter* 21(39), 395502 (2009)

49. P. Giannozzi, et al., Advanced capabilities for materials modelling with Quantum ESPRESSO. *J. Phys.: Condens. Matter* 29(46), 465901 (2017)

50. J. P. Perdew; K. Burke; M. Ernzerhof, Generalized Gradient Approximation Made Simple. *Phys. Rev. Lett.* 77(18), 3865-3868 (1996)

51. H. J. Monkhorst; J. D. Pack, Special points for Brillouin-zone integrations. *Phys. Rev. B* 13(12), 5188-5192 (1976)

52. J. Heyd; G. E. Scuseria; M. Ernzerhof, Hybrid functionals based on a screened Coulomb potential. *J. Chem. Phys.* 118(18), 8207-8215 (2003)

53. A. Togo; F. Oba; I. Tanaka, First-principles calculations of the ferroelastic transition between rutile-type and CaCl2-type SiO2 at high pressures. *Phys. Rev. B* 78(13), 134106 (2008)

54. Q. Wu; S. Zhang; H.-F. Song; M. Troyer; A. A. Soluyanov, WannierTools: An open-source software package for novel topological materials. *Comput. Phys. Commun.* 224(405-416 (2018)

55. G. Pizzi, et al., Wannier90 as a community code: new features and applications. *J. Phys.: Condens. Matter* 32(16), 165902 (2020)